\documentclass[prd,twocolumn,a4paper,showkeys,showpacs,nofootinbib]{revtex4-1}

\usepackage[frenchb,english]{babel}
\usepackage[latin1]{inputenc}
\usepackage{enumerate}
\usepackage{amsmath}
\usepackage{amsthm}
\usepackage{amssymb}
\usepackage{graphicx}
\usepackage{floatflt}
\usepackage{fancybox}
\usepackage{stmaryrd}
\usepackage{appendix} 
\usepackage{enumitem} 

\usepackage{hyperref} 

\usepackage{color}

\newcommand{\Ai}{\textrm{Ai}}
\newcommand{\AI}{\textrm{PAi}} 

\newcommand{\om}{\omega}
\newcommand{\Om}{\Omega}

\newcommand{\p}{\partial}

\newcommand{\tp}{{\rm tp}}
\newcommand{\br}{{\rm br}}
\newcommand{\li}{{\rm lin}}
\newcommand{\Lam}{\Lambda}

\newcommand{\be} {\begin{equation}}
\newcommand{\ee} {\end{equation}}
\newcommand{\bsub}{\begin{subequations}}
\newcommand{\esub}{\end{subequations}}
\newcommand{\bea}{\begin{eqnarray}}
\newcommand{\eea}{\end{eqnarray}}
\newcommand{\bi} {\begin{itemize}}
\newcommand{\ei} {\end{itemize}}
\newcommand{\ben} {\begin{enumerate}}
\newcommand{\een} {\end{enumerate}}
\newcommand{\bmat} {\begin{pmatrix}}
\newcommand{\emat} {\end{pmatrix}} 
\newcommand{\bal} {\begin{aligned}}
\newcommand{\eal} {\end{aligned}}
\newcommand{\btab}{\begin{tabular}}
\newcommand{\etab}{\end{tabular}}

\newcommand{\eq}[1]{Eq.~\eqref{#1}}

\begin{document}
\selectlanguage{english}

\title{Hawking radiation with dispersion: The broadened horizon paradigm}

\author{Antonin Coutant}
\email{antonin.coutant@aei.mpg.de}
\affiliation{Max Planck Institute for Gravitational Physics, Albert Einstein Institute, Am Muhlenberg 1, 14476 Golm, Germany}

\author{Renaud Parentani}
\email{renaud.parentani@th.u-psud.fr}
\affiliation{Laboratoire de Physique Th\'eorique, CNRS UMR 8627, B\^atiment 210, Universit\'e Paris-Sud 11, 91405 Orsay Cedex, France}

\date{\today}

\begin{abstract}

To identify what replaces the key notion of black hole horizon when working with theories which break Lorentz invariance at high energy, 
we study the modes responsible for the Hawking effect in the presence of high frequency dispersion. 
We show that they are regularized across the horizon over a short length 
which only depends on the scale of dispersion and the surface gravity. Moreover, outside this width, 
short and long wavelength modes no longer mix. These results can be used to show that the spectrum is hardly modified by dispersion 
as long as the background geometry does not vary significantly over this length. 
For relevant frequencies, the regularization differs from the usual WKB resolution of wave singularity near a turning point.

\end{abstract}

\pacs{04.62.+v,    
04.70.Dy,   
11.30.Cp   
}
\keywords{Lorentz violating theories, Hawking effect, Black hole thermodynamics, Analog gravity} 

\maketitle

\section{Introduction}
The invariance under the Lorentz symmetry group is central to our current description of high energy processes~\cite{Weinberg1}.
Yet, being noncompact, we do not know if this group is an {\it exact} symmetry of nature~\cite{Mattingly05}. 
This inherent incompleteness of observational data is reinforced by our lack of knowledge about the ultraviolet (UV) structure of quantum gravity.  
It is therefore of value to develop alternative approaches where the Lorentz group is violated in the UV,  
and examine what are the consequences. As we shall see, 
Hawking radiation can play a crucial role in revealing them. At present, the laws of black hole thermodynamics are 
poorly understood~\cite{Jacobson08b,Busch12,Berglund12} in Lorentz violating theories (LVT), as Einstein-Aether~\cite{Jacobson01} or Ho\u rava-Lifshitz gravity~\cite{Horava09}. 
In fact the Hawking process itself is not yet understood~\cite{Cropp13}. 
Moreover, modern discussions about
black hole evaporation, such as the ``firewall'' proposal~\cite{Marolf13}, also
heavily rely on assumptions concerning the UV behavior of the theory.
The present paper aims at revealing the properties of the modes in the vicinity of the horizon that are  
specific to LVT. Interestingly, these properties share some similarities with those of 
attempts to take into account gravitational effects neglected 
in the semi-classical scenario~\cite{York83,Casher96,Parentani07}. 
Moreover they also apply to condensed matter systems in the context of analog gravity~\cite{Unruh81,Balbinot06,Barcelo05}, 
where fluid flows are used to mimic black hole geometries, 
and to test the Hawking process in the presence of UV dispersion, 
see e.g.~\cite{Weinfurtner10,Michel14}. 

In addition to the space-time metric, LVT are endowed with a 
dynamical vector field $u^\mu$ that introduces a preferred frame, which is 
used to covariantly implement the physical processes breaking the Lorentz symmetry~\cite{Jacobson01,Kostelecky03}. 
In this paper, for simplicity, the vector field shall be taken geodesic (freely falling), and only  
high energy dispersion shall be considered. First deviations with respect to relativity can be described by 
\be
\Om^2 = F^2(p) = p^2 ( 1- p^2/\Lambda^2) 
, \label{disprel}
\ee
where $\Om = -u^\mu P_\mu$ is the frequency measured in the preferred frame, $P^\mu$ the four-momentum 
of the particle, and $p^2$ the squared norm of its spatial momentum perpendicular to $u^\mu$~\cite{Jacobson96}. 
The minus sign in \eq{disprel} means that the dispersion is subluminal. (The superluminal case can be treated in a similar way, see Sec.III.E in~\cite{Coutant11}). 
$\Lam$ defines the UV cutoff above which Lorentz invariance ceases to be valid.~\footnote{In the firewall debates, 
a (local) characterization of the validity domain of the effective (Lorentz invariant) field theory, which is 
``set by some UV cutoff''~\cite{Almheiri13b}, requires to adopt a preferred frame.}

We now recall why the Hawking process acts as a microscope probing ultrahigh energy physics. 
For relativistic fields, the stationary modes $\phi_\om$ responsible for the Hawking effect 
are singular on the horizon: for $|x| \to 0$, one finds 
\be
\phi_\om \propto |x|^{i\om/\kappa}, \label{relat}
\ee
where $\omega/\kappa$ is the ratio of their Killing frequency over the surface gravity, and where $x$ is the proper distance from the horizon measured in 
the preferred frame. Importantly, the singular behavior of \eq{relat} unambiguously fixes the temperature of the emitted radiation~\cite{Unruh76}. 
Indeed, the regularity of the state across the horizon fixes the ratio of the coefficients weighing $\phi_\om$ on either side of $x=0$, and this in turn fixes the temperature to be the Hawking one: $T_H = \kappa/2\pi$ in units where $c = \hbar = k_B = 1$. From \eq{relat}, one can see that $\Om \sim p \sim {\om/ \kappa x}$ increases without bound as $x \to 0$. This blueshift connects the low energy physics $\Om \sim \kappa$, to the UV 
physics where the standard description based on free relativistic fields propagating in classical background might break down. 
This raises the {\it trans-Planckian question}~\cite{Jacobson91,Jacobson99}, 
namely, to what extent the predictions derived from \eq{relat} actually depends on the (unknown) UV 
behavior of the theory. This question played a crucial role in the development of LVT. 
It is now clear that in the vicinity of the horizon, the field propagation is highly sensitive to a modification such as that of \eq{disprel}.

Since~\cite{Unruh95}, attention has been mainly given to the modifications of the asymptotic  
spectrum due to high momentum dispersion~\cite{Brout95,Corley97,Unruh04,Macher09,Coutant11,Finazzi10b,Finazzi12}. 
In this paper instead, we consider the near horizon properties of the modes. 
Interestingly, we shall see that: first, the dispersive modes  
involve a single, composite, and $\om$-independent short length scale, and second, they 
display two very distinct behaviors depending on the value of $\om/\kappa$. 

\section{Settings}
As \eq{relat} is found irrespectively of the mass and the orbital momentum, 
we shall work with massless $1+1$ dimensional fields. 
The stationary geometry shall be described by the line element 
\be
ds^2 = dt^2 - (dx - v(x)dt)^2 , \label{PG} 
\ee 
where $v < 0$ and where $dt = u_\mu dx^\mu$ is the freely falling proper time. 
The event horizon is located at $v^2 = 1$, and the interior of the black hole, $|v| > 1$,
is here $x<0$. 
We work with a massless field propagating in \eq{PG} and obeying \eq{disprel}. 
At fixed $\om$, 
the mode $\phi_\om$ obeys~\cite{Unruh95,Brout95} 
\be
[(\om + i\p_x v)(\om + iv\p_x) - F^2(-i\p_x) ] \phi_\om = 0. \label{modeeq}
\ee
Close to the horizon, one has $v \sim - 1 + \kappa x$. This approximation is valid only for a finite range of $x$, that we call $x_\li$. 
In usual black hole geometries, $x_\li \lesssim 1/\kappa$. 

\begin{figure}[!ht]
\begin{center} 
\includegraphics[width=\columnwidth]{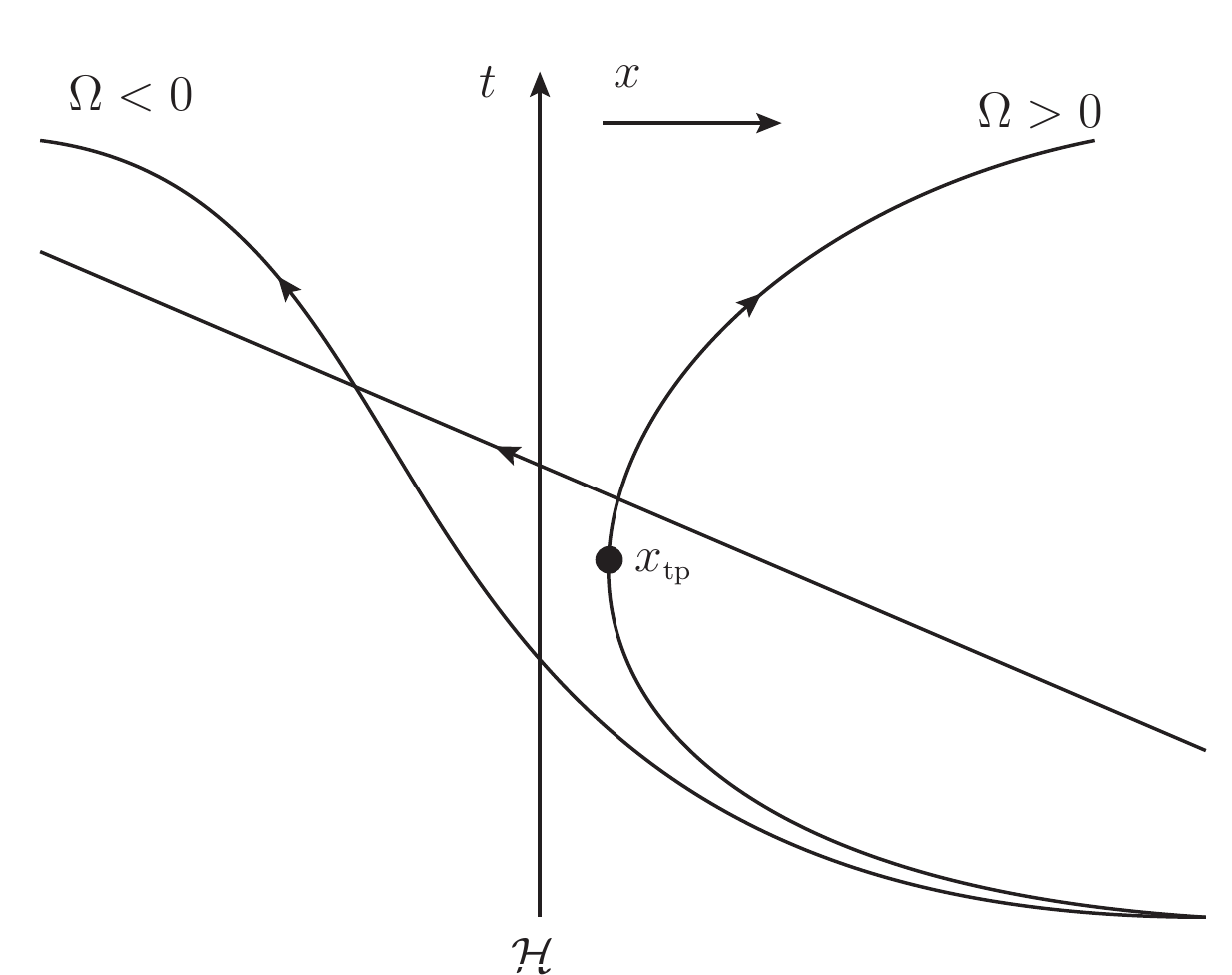}
\end{center}
\caption{Space-time structure of characteristics in the near horizon region. 
The straight line is the characteristic of the spectator mode which plays no role in the Hawking effect. 
The curved characteristics correspond to the modes for $\om >0$ with positive (right side) and negative (left) freely falling frequency $\Om$. 
For a detailed description, we refer to Sec.I.D of~\cite{Coutant11}. 
}
\label{BH_traj_fig} 
\end{figure}

The first manifestations of dispersion show up in the characteristics of \eq{modeeq}. 
When considered backwards in time, instead of focusing on the horizon as in the relativistic case, they  
are swept away at short wavelengths, see Fig ~\ref{BH_traj_fig}.
Then, if $\Om = \om - v p < 0$, the trajectory crosses the horizon and falls into the hole, but if $\Om > 0$, it bounces back at a finite distance of the horizon. 
The turning point occurs at $p_\tp = \om^{1/3} \Lam^{2/3}$ in momentum space, 
and is localized at 
\be
x_\tp(\om) =\frac{3\om}{2\kappa p_\tp} = \frac{3}{2\kappa} \left(\frac{\om}{\Lam} \right)^{2/3}.
\label{xtp}
\ee
This expression is valid if $x_\tp$ lies within the near horizon region, i.e., $x_\tp \ll x_\li$. 
\eq{xtp} gives the first composite length of the problem. 

When considering the wave equation \eqref{modeeq}, its resolution turns out to be simpler in $p$-space. 
In fact, using $\hat v = -1 + i\kappa \p_p$, the solution of \eq{modeeq} neatly factorizes~\footnote{This comes from the fact that $v = - 1 + \kappa x$ considered globally corresponds to de Sitter space, 
which possesses an extra symmetry compatible with $u^\mu$, see~\cite{Busch12} for details.} 
as $\tilde \phi_\om(p) =  p^{-i \frac{\om}\kappa - 1} \times \chi(p) e^{-i \frac{p}\kappa}$~\cite{Brout95,Jacobson07b}. 
The first factor is the relativistic mode, since the Fourier transform of \eq{relat} gives  
$|p|^{- i\om/\kappa -1}$. The function $\chi$ obeys $-\kappa^2 p^2 \p_p^2 \chi = F^2(p) \chi$, 
which is $\om$-independent. To solve this equation, we use the WKB approximation in $p$-space. 
As explained in~\cite{Balbinot06,Coutant11}, this amounts to neglecting the mixing of Hawking modes with left movers. 
This is valid when $\kappa/\Lam \ll 1$, which we assume to be satisfied. 
We also consider $\om/\Lambda \ll 1$ because it allows us 
to work in a {\it weakly dispersive regime}, where $F \sim p - p^3/(2\Lam^2)$ in the phase of $\tilde \phi_\om$, 
and $F \sim p$ in its slowly varying amplitude (see~\cite{Coutant11} for more details about this). 
We thus obtain
\be
\tilde \phi_\om(p) = \frac{p^{-i \frac{\om}\kappa - 1}}{\sqrt{2\pi}} \exp \left(- i \frac{p^3}{6\Lam^2 \kappa} \right). \label{pWKB}
\ee
From the exponential factor, we see that the stationary modes  
involve a second composite length,  
\be
d_\br = \frac1{(2\kappa)^{1/3} \Lam^{2/3}}.
\label{dbr} 
\ee
As we shall see below, $d_\br$ plays a prevalent role with respect to 
the $\om$-dependent length of \eq{xtp}.

\section{Near horizon modes} 
To get the spatial properties of the mode, we inverse Fourier transform \eq{pWKB}:
\be
\phi_\om^{\mathcal C}(x) = 
\int_{\mathcal C} \exp{ \left[i \left( q \frac{x}{d_\br} - \frac\om\kappa \ln(q) - \frac{1}{3} q^3 \right) \right]} \frac{dq}{2\pi q}.
\label{inverseFT}
\ee
We introduced the adimensionalyzed wave vector $q \doteq p d_\br$. 
The various solutions of \eq{modeeq} are recovered by adopting 
different contours $\mathcal C$ in the complex $p$ plane~\cite{Brout95,Corley97,Unruh04,Coutant11,Fleurov12}. 
Irrespectively of the contour $\mathcal C$ and the value of $\om$, we see that 
$\phi_\om^{\mathcal C}$ only depends on $z \doteq x/d_\br$.
As we shall now see, the integral representation of \eq{inverseFT} contains all the relevant information, 
i.e., both the mode profiles in the near horizon region\footnote{Note that $\phi_\om^{\mathcal C}$ of \eq{inverseFT} solves 
\be
\p_z^3 \phi_\om + z\p_z \phi_\om - i\frac{\om}{\kappa} \phi_\om = 0, \nonumber 
\ee
and is thus a linear combination of hypergeometric functions ${}_1 F_2$~\cite{AbramoSteg}, as noticed in~\cite{Fleurov12}.
From the 3 independent solutions, only 2 are physical, since the third combination grows without bound on the left side of the horizon. 
Because the identification of the physical modes is very cumbersome, 
the mode analysis in the near horizon cannot be done in a transparent manner in terms of ${}_1 F_2$.}, and
the Bogoliubov coefficients encoding the Hawking effect. 

In the following, we analyze the mode that decays inside the horizon. It is proportional to the 
outgoing mode $\phi_\om^{\rm out}$. 
(A similar analysis, leading to the same conclusions, can be done for the partner mode, orthogonal to $\phi_\om^{\rm out}$.) 
To obtain it, the branch cut of the $\ln (q)$ should be put on $i \mathbb R^+$. 
The large $z$ expansion of \eq{inverseFT} then gives~\cite{Coutant11} 
\be \bal
\phi_\om^{\mathcal C}&(x) \sim e^{-i\frac{\pi}4} \frac{e^{i \frac23 z^{3/2}} e^{-i\frac{\omega}{2\kappa} \ln z }}{\sqrt{4\pi z^{3/2}}} \\
&- \frac{\beta_\om}{\alpha_\om} e^{i\frac{\pi}4} \frac{e^{-i \frac23 z ^{3/2}} e^{-i\frac{\omega}{2\kappa} \ln z }}{\sqrt{4\pi z^{3/2}}} + \frac1{\alpha_\om} z^{i\frac{\om}{\kappa}} \sqrt\frac{\kappa}{2\pi\om} .
\eal \label{Connection} 
\ee
This equation means that $\phi_\om^{\mathcal C}$ reduces to a sum of WKB waves that no longer mix. 
When the latter have unit norms~\cite{Coutant11}, the coefficients governing their respective weight define the near horizon scattering coefficients~\footnote{This mode mixing will be in general completed by some extra scattering taking place further away from the horizon~\cite{Coutant12}. For lower values of $p$, on the outside region, this gives rise to greybody factors~\cite{Page76}. 
In LVT, some extra scattering could also occur at higher values of $p$. As a result, the incoming modes  
could not arrive in their ground state, thereby stimulating the Hawking process. 
For a subluminal dispersion, this possibility is suppressed by the adiabatic propagation from infinity to the horizon. 
For a superluminal dispersion, as in Ho\u rava-Lifshitz gravity~\cite{Cropp13}, 
the mode scattering on the universal horizon may instead significantly affect the resulting spectrum.}.
Taking into account that the last term of \eq{Connection} describes the low momentum outgoing mode, 
$\alpha_\om$ and $\beta_\om$ can be shown to be the Bogoliubov coefficients encoding
the Hawking effect~\cite{Coutant11}. Their ratio here obeys  
\be
\beta_\om = e^{- \frac{\pi \om}{\kappa}} \alpha_\om, \label{BogoHR}
\ee
as in the relativistic case. 
Thus, the temperature is still given by the standard expression $T_H = \kappa/2\pi$. 
On the other side of the horizon ($z<0$), the mode decays as $\sim e^{- \frac23 |z|^{3/2}}$. 

We now study the validity range of these results. A careful computation~\cite{Coutant11} shows that the corrections are negligible when 
\bsub \label {VC} \bea 
 |z| &\gg& 1 , \label{VCdbr} \\
|z| &\gg& \om/\kappa . \label{VCxtp} 
\eea \esub 
However, they also require $d_\br |z| \lesssim x_\li$, since $v \sim - 1 + \kappa x $ has been used. 
When $\Lam/\kappa$ is large enough, the spatial range satisfying these three 
inequalities is quite large. In the frequency range relevant for the Hawking effect, i.e. for $\om \lesssim \kappa$, 
\eq{VCdbr} implies \eq{VCxtp}. Instead, for $\om \gg \kappa$, 
the Hawking process is then exponentially suppressed. 
Therefore, \eqref{VCdbr} is the most relevant condition to obtain \eq{Connection}. 
This shows that the mode mixing responsible for the Hawking effect occurs in a region around the horizon of size $d_\br$. 
Additionally, the modes can only resolve the precise location of the turning point from the horizon when $x_\tp \gg d_\br$, which corresponds to $\om \gg \kappa$ and the suppression of the Hawking effect. This shows that the relevant length scale that characterizes the Hawking process is $d_\br$, and not $x_\tp$. 
Because of Eqs.~\eqref{VC}, \eq{Connection} says nothing about the mode behavior 
in a close vicinity of the horizon. 
In what follows, to characterize this behavior, we separately analyze \eq{inverseFT} for low and high frequency. 

\subsection{Small frequency regime, $\om \lesssim  T_H$} 
With the branch cut of $\ln (q)$  on $i \mathbb R^+$, 
the $\om \to 0$ limit of \eq{inverseFT} is proportional to the {\it primitive integral} 
of the Airy function $\Ai(-z)$ that vanishes for $z \to -\infty$, see ~\cite{AbramoSteg}. 
Calling it $\AI(-z)$, we have 
\be
\phi_0^{\mathcal C}(x) = i \AI \left(- z\right). \label{AiUndul}
\ee
This result is consistent with \eq{Connection}, as one sees by comparing it to the asymptotics 
of $\AI$. In white hole flows, this mode gives the spatial profile of the {\it undulation} 
studied in~\cite{Coutant13,Mayoral11}, and observed in~\cite{Weinfurtner10}. 

\begin{figure}[!ht]
\begin{center} 
\includegraphics[width=\columnwidth]{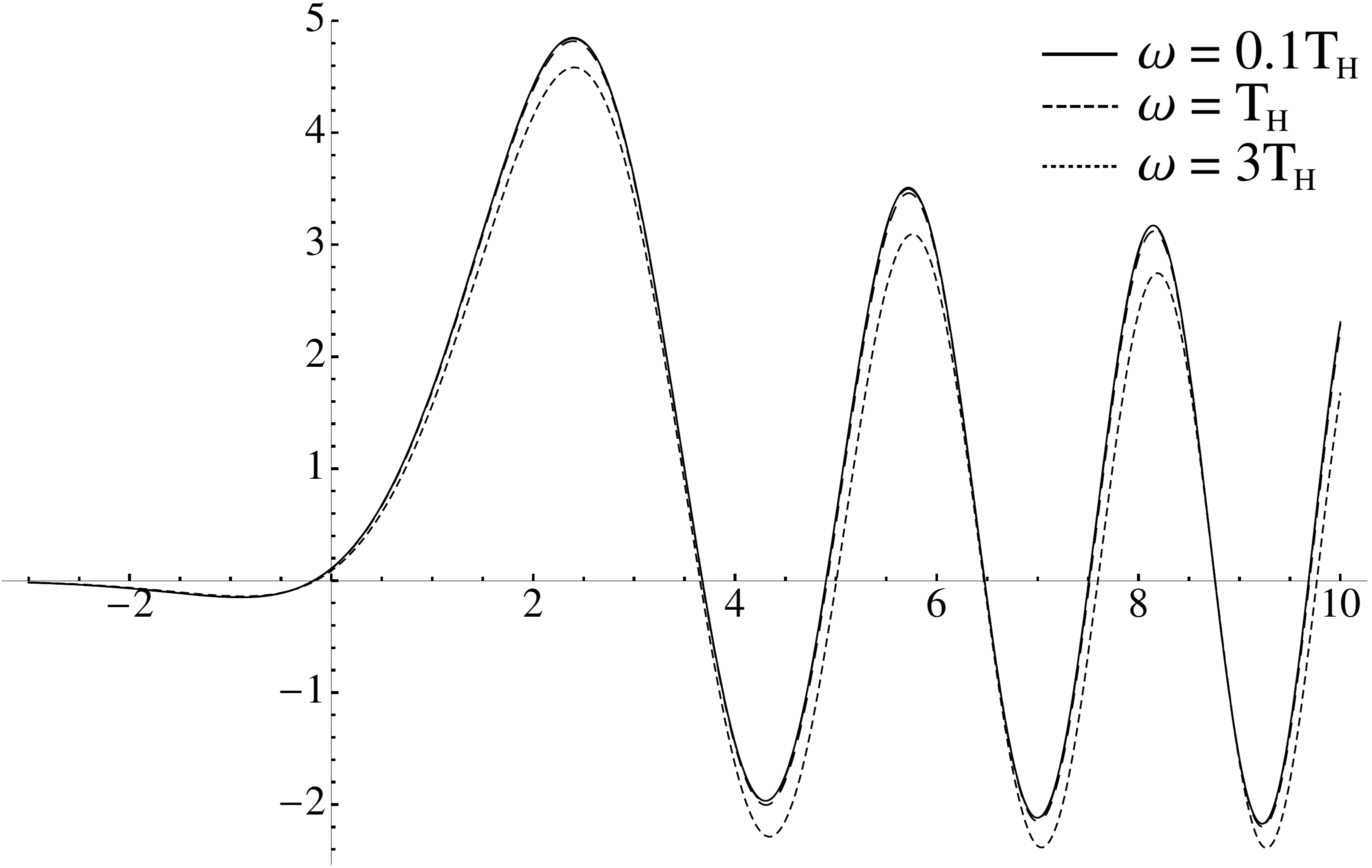}
\end{center}
\caption{Plot of $-\p_z \psi_\om(z)$ 
of \eq{AI_ReIm} as a function of $z = x/d_\br$, and three values of $\om$. For numerical reasons, we plotted the derivative instead of $\psi_\om$ itself. 
Only the curve for $\om = 3T_H$ (dashed line) can be distinguished from the others. This establishes 
that \eq{AI_ReIm} offers an accurate description of the near horizon profile 
for {several} $d_\br$ lengths, and for $0 \leqslant \om \lesssim 3T_H$. 
} 
\label{small_om_fig} 
\end{figure}

For $0 < \om \lesssim T_H$, \eq{Connection} predicts a modulation of \eqref{AiUndul} 
by $\exp(i \om\ln z /2\kappa )$ for $z \gtrsim 1$. The location of the first node is given by $x_{\rm zero}/d_\br  \sim  e^{4\pi \kappa / \om}$. For $\om \sim T_H$, 
we thus have $x_{\rm zero}/d_\br \sim e^{8\pi^2}$. Hence, this modulation
possesses a wavelength much larger than $d_\br$, possibly even larger than the near horizon size $x_\li$. 
It is thus a subdominant effect, barely visible as long as $\om \lesssim T_H$. 
As a result, the first significant effect comes from $\beta_\om / \alpha_\om = e^{- \pi \om/\kappa} \neq 1$. 
To establish this, we decompose the mode as 
\be
\phi_\om^{\mathcal C}(x) = i \frac{1+e^{-\frac{\pi \om}\kappa}}{2} 
\varphi_\om(z) + \frac{1-e^{-\frac{\pi \om}\kappa}}{2} \psi_\om\left(z\right), 
\label{AI_ReIm}
\ee 
where $\varphi_\om$ and $\psi_\om$ are {\it real} functions. 
Once having factorized the two prefactors which arise from $\beta_\om$ and $\alpha_\om$, 
the residual dependence in $\om$ of $\varphi_\om$ and $\psi_\om$ is minimized, and no longer significant. 
As a result, $\varphi_\om$ can be replaced by $\phi_0^{\mathcal C}$ 
of \eq{AiUndul}. Similarly, $\psi_\om$ is also essentially independent of $\om$. 
This is neatly confirmed in Fig.\ref{small_om_fig}. 

In conclusion, Eqs.~\eqref{AiUndul}, \eqref{AI_ReIm}, and Fig.~\ref{small_om_fig} 
explicitly give the near horizon properties of the dispersive mode  $\phi_\om^{\rm out}$
for {\it several}  $d_\br$ lengths,
and for frequencies $0 \leqslant \om \lesssim 3 T_H$, which is the most relevant domain for 
the Hawking effect. 
This is our principal result. 

\subsection{Large frequency regime $\om \gg T_H$} 
\begin{figure}[!ht]
\begin{center} 
\includegraphics[width=\columnwidth]{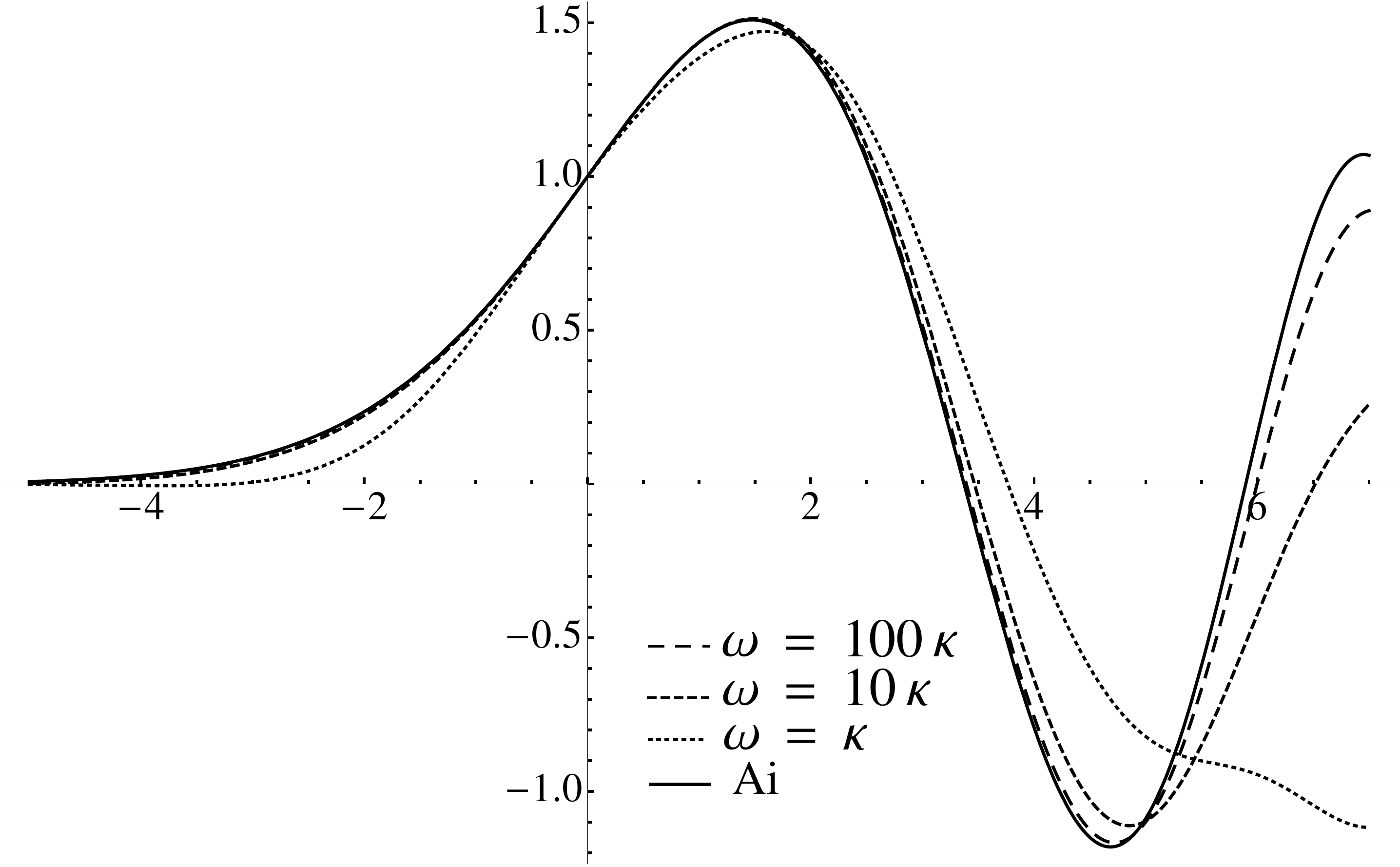}
\end{center}
\caption{Plot of $e^{- i z q_\tp} \times \phi_\om^{\mathcal C}(x_\tp + z d_\br)$ as a function of $z$, 
various values of $\om$, and also compared with the Airy function $\Ai(-z/3^{1/3})$. 
The different curves are normalized to 1 at the turning point $z_\tp(\om)$, which is here set at $z = 0$. 
By numerically comparing the values at the first peak, 
we found that the error decreases as $\sim (\kappa/\om)^\gamma$ with the exponent $1/3.25 \lesssim \gamma \lesssim 1/3.15$, in agreement  
with \eq{RAiry_validity} to a good accuracy. Since $q_\tp \gg 1$, 
\eq{phitp} is valid for many short wavelength oscillations due to $e^{i z q_\tp}$. 
}
\label{large_om_fig} 
\end{figure}

When $\om$ is larger than $T_H$, the $\beta_\om$-term in \eq{Connection} is exponentially small. 
Hence, one is left with a total reflection. 
To obtain the mode near the turning point, we now follow the standard procedure. 
It consists in expanding the phase of the integrand of \eq{inverseFT}, i.e., 
$W(z,q) = z q - \frac{\om}\kappa \ln(q) - \frac{1}{3} q^3$, to third order in $\Delta q = q - q_{\rm tp}$, 
where $q_\tp(\om) = d_\br p_\tp$. Performing the $q$-integration, by construction, one obtains an Airy function: 
\be
\phi_\om^{\mathcal C}(x) = \frac{e^{i\theta_\tp}}{3^{1/3} q_\tp } \times e^{i z q_\tp} \times 
 \Ai\left(-\frac{z-z_\tp}{3^{1/3}} \right), \label{phitp}
\ee
where $z_\tp = x_\tp/d_\br$, see \eq{xtp}, and where $\theta_\tp = -\om/(3\kappa) \ln(\om/2\kappa) - \om/(6\kappa)$. 
\eq{phitp} is valid when 
\be
\left|\frac{\kappa}{\om}\right|^{1/3} \ll 1,
\label{RAiry_validity}
\ee 
and when $|x-x_\tp| \ll x_\tp$. These conditions are supported by Fig.~\ref{large_om_fig}. 
Hence, we see that the usual WKB resolution becomes valid at high frequencies, precisely when Hawking radiation fades away.

\section{Smoothing out short-distance details} 
When computing the Hawking spectrum in the presence of dispersion, 
it is {\it a priori} tempting to take into account the $\om$-dependence of \eq{xtp}, and 
to use the value of the gradient $\kappa(x) = \partial_x v$ 
evaluated at $x_\tp(\om)$ in the place of the surface gravity $\kappa$. 
Yet, no such dependence was found in numerical analysis of the spectrum~\cite{Corley96,Macher09,Finazzi10b}. 

To clarify these observations, as in~\cite{Finazzi10b}, we consider background profiles of the form 
$v = v_0(x) + \delta v(x)$, where $v_0$ is smooth enough so that the above analysis applies, 
and where $\delta v$ is a small perturbation. 
If $\delta v \ll 1$,  adapting the {\it distorted wave Born approximation}~\cite{Newton} to mode amplification, 
the induced correction of the $\beta$ Bogoliubov coefficient is 
\be
\delta \beta_\om = 2i\pi \int \left[ \phi_{-\om}^{\rm out} \p_x(\delta v \, \pi_\om^{\rm in}) + \pi_{-\om}^{\rm out} \delta v\, \p_x \phi_\om^{\rm in} \right] dx, 
\label{DWBA}
\ee 
where $\pi(x)= (\p_t + v\p_x)\phi$ is the momentum conjugated to $\phi$, and where $\phi_\om^{\rm in}$ ($\phi_\om^{\rm out}$) is the incoming (outgoing) positive norm mode propagating in the unperturbed flow. From this expression, we clearly see that if the scale of variation of $\delta v$ is much shorter than $d_\br$, the integration washes out $\delta v$, and $\delta \beta_\om$ essentially vanishes. 
This establishes that the finite resolution of the modes erases the details of the background on scales smaller than $d_\br$. Therefore, instead of $\kappa(x_\tp)$, the effective surface gravity should be obtained by averaging $\partial_x v$ over a broadening length, 
as discussed in~\cite{Finazzi10b}.

\section{Conclusions} 
Our results show that, in LVT with quartic dispersion, Hawking radiation can be understood by interpreting the horizon as broadened over a length $d_\br = (2\kappa \Lam^2)^{-1/3}$. 
Firstly, the mode mixing responsible for the Hawking effect now occurs within a region of size $d_\br$ across the horizon. 
Secondly, when $d_\br \ll x_\li$, i.e., when the local surface gravity does not change over $d_\br$,
the standard Hawking spectrum is recovered. 
Thirdly, since the modes are regulated over $d_\br$, 
details of the near horizon geometry much smaller than $d_\br$ are washed out. 
For a black hole of mass $M$, $d_\br \propto M^{1/3}$ in Planck units. 
Interestingly, the same scaling was found by studying horizon fluctuations in~\cite{Casher96}. 
These results could also be tested in future analogue gravity experiments based, e.g., on surface waves in flumes~\cite{Weinfurtner10,Euve14}. 
We believe they should also play a key role in the (not yet understood) black hole thermodynamics in LVT.

\acknowledgements{We thank Xavier Busch for help with the numerical part of this work.}

\bibliographystyle{utphys}
\bibliography{Biblio}

\end{document}